\documentclass[a4paper,11pt]{article}
\usepackage{aaskaiid}
\usepackage{orcidlink}

\graphicspath{{./figures/}}

\title{From Atomic Gas to Star Formation}
\ShortTitle{\textsc{H\,i} and Star Formation}

\author[1]{Erik Rosolowsky\orcidlink{0000-0002-5204-2259}}
\ShortName{Rosolowsky et al.} 
\author[2]{Eric Koch\orcidlink{0000-0001-9605-780X}}
\author[3]{Sambit Roychowdhury\orcidlink{0000-0002-5820-4589}}
\author[4]{Luca Cortese\orcidlink{0000-0002-7422-9823}}
\author[5,6]{Filippo M. Maccagni\orcidlink{0000-0002-9930-1844}}
\author[4]{Barbara Catinella\orcidlink{0000-0002-7625-562X}}
\author[7]{Nushkia Chamba\orcidlink{0000-0002-1598-5995}}
\author[8]{Timothy A. Davis\orcidlink{0000-0003-4932-9379}}
\author[2]{Cosima Eibensteiner\orcidlink{0000-0002-1185-2810}}
\author[2]{Eric Murphy\orcidlink{0000-0001-7089-7325}}
\author[9]{Mamta Pandey-Pommier\orcidlink{0000-0001-5829-1099}}
\author[10]{Amidou Sorgho\orcidlink{0000-0002-5233-8260}}

\affiliation[1]{Department of Physics, University of Alberta, 4-183 CCIS, Edmonton, Alberta, Canada}
\emailAdd{rosolowsky@ualberta.ca}
\affiliation[2]{National Radio Astronomy Observatory, 520 Edgemont Road, Charlottesville, VA 22903, USA}
\emailAdd{koch.eric.w@gmail.com}
\affiliation[3]{University Observatory Munich (USM), Ludwig Maximilian University of Munich (LMU), Scheinerstraße 1, 81679 Munich, Germany}
\emailAdd{sambit.roychowdhury@physik.lmu.de}
\affiliation[4]{International Centre for Radio Astronomy Research, The University of Western Australia, 35 Stirling Highway, Crawley, Perth, WA 6009, Australia}
\affiliation[5]{INAF -- Osservatorio Astronomico di Cagliari, via della Scienza 5, 09047, Selargius (CA), Italy}
\affiliation[6]{Wits Centre for Astrophysics, School of Physics, University of the Witwatersrand, 1 Jan Smuts Avenue, 2000, Johannesburg, South Africa}
\emailAdd{filippo.maccagni@inaf.it}
\affiliation[7]{NASA Ames Research Center, Moffett Field, CA 94035, USA}
\affiliation[8]{Cardiff Hub for Astrophysics Research \&\ Technology, School of Physics \&\ Astronomy, Cardiff University, Queens Buildings, Cardiff, CF24 3AA, UK}
\affiliation[9]{Pole Scientific, University Catholic of Lyon, Campus Saint-Paul, 10 place des Archives 69288, Lyon Cedex 02, France}
\emailAdd{mamtapommier@gmail.com}
\affiliation[10]{Instituto de Astrof\'isica de Andaluc\'ia (IAA-CSIC), Glorieta de la Astronom\'ia s/n, E-18008 Granada, Spain}

\abstract{Since the 2015 SKA Chapters, the advent of panchromatic observations of nearby galaxies—capable of resolving the interstellar medium (ISM) into individual star-forming regions and molecular clouds—has revolutionized our understanding of the star formation cycle in galaxies.  Despite these advances, our understanding of the gas cycle in galaxies remains  limited because we know little about how the cold atomic phase of the ISM, which dominates its mass budget, participates in the cycle. Specifically, the processes regulating the condensation of atomic gas into molecular gas, the balance between warm and cold atomic gas, and their dependence on local environment and star formation activity have been explored only in the Milky Way and the Magellanic Clouds.

The SKA presents a new opportunity to observe the atomic ISM within nearby galaxies at comparable resolution to other facilities and is the only facility that can make wide-area surveys of atomic gas across a statistical sample of galaxies at high ($<100$~pc) spatial resolution.  Such observations will directly address the open questions about the evolution of the star-forming ISM. We present recent results from SKA-like observations of the Local Group made with SKA pathfinders and the Jansky VLA to illustrate the promise of replicating those observations for hundreds of nearby galaxies, reaching out to the Virgo cluster.
}

\begin{document}
\maketitle

\section{Introduction}

The internal processes of galaxy evolution are shaped by the matter cycle. This cycle encompasses how baryonic matter is accreted into galaxies, changes phases in the interstellar medium (ISM), and forms into stars. In this entire cycle, the key interface is the connection between the ISM of a galaxy and the stars, mediated through star formation from the gas in the ISM and then the consequent stellar and AGN feedback that reshapes and redistributes the ISM.

Since the publication of the first \textit{Advancing Astrophysics with the SKA} proceedings \citep{bourke2015advancing}, we have made great advances in understanding the matter cycle through resolved studies of nearby galaxies.  These have been driven by substantial investments in mapping out the gas and stars using facilities like HST, JWST, ALMA, and the VLT/MUSE. These combined efforts have provided a well resolved ($<100$ pc) view of nearly all major mass phases in galaxies: stars, molecular gas, dust, ionized gas, even hot plasma. \citep[e.g., the PHANGS surveys,][]{phangs_alma}.  The community has developed a rich set of multi-wavelength data combined with a sample size that is sufficiently large to resolve the underlying physical trends from the happenstance that shapes the evolution of a single galaxy. 

Since star formation at $z=0$ happens in the molecular ISM, the study of star formation has focused on the properties of that molecular gas.  Thanks to ALMA's transformational capability to make {10-100 pc}-resolution maps of CO in nearby galaxies using only short periods of time, the past decade has seen observations of hundreds of galaxies developing a suite of highly-resolved gas maps.  By pairing these resolved gas maps with high resolution tracers of star formation (H$\alpha$ and other recombination line studies, far ultraviolet emission, and short wavelength radio observations), we have been able to characterize the structure of molecular gas and calibrate how efficiently molecular gas forms stars.  \citet{schinnerer24} review recent progress in this field, highlighting that molecular gas properties are shaped by the broader galactic environment and that molecular gas appears to form stars rapidly and be destroyed efficiently by stellar feedback on few Myr timescales.  Stellar feedback plays a central role in regulating this process, establishing a vertical dynamical equilibrium in disks \citep{sun_2020}.  Finally, the star formation process is found to be globally inefficient with a star formation efficiency per free fall time of $\epsilon_\mathrm{ff} \sim 0.5\%$.  

This understanding of star formation is fairly simple. Star formation in a galaxy has a tight correlation with the amount of the molecular gas  \citep{leroy13,delosreyes19,kennicutt21}. 
The local star formation rate volume density seems to follow a model where $\dot{\rho}_\star = \epsilon_\mathrm{ff} \rho_\mathrm{mol}/t_\mathrm{ff}$, where $\rho_\mathrm{mol}$ is the volume mass density of molecular gas and $t_\mathrm{ff}$ is the free-fall time.  Star formation is inefficient with $\epsilon_\mathrm{ff}\ll 1$ with only modest sensitivity to local gas conditions \citep[e.g.,][]{km05,federrath12,padoan12,burkhart18}.  Observations on 100-pc scales show that $\epsilon_\mathrm{ff}$ has a small value near what is argued for theory, and the data show some variation with galactic environment \citep{leroy_2025}, though these scalings are not readily explained by contemporary theories.

Because it is actively forming stars, the molecular phase appears to be short-lived \citep{schinnerer24} with the molecular gas rapidly disrupted and converted to other phases of the ISM through stellar feedback.  This means that the key step for studying the star formation process has become the study of the \textit{formation} of the molecular ISM from the cold atomic medium. At $z=0$, most of the mass of the ISM is in the atomic phase \citep{saintoge22}, which is thought to remain true at higher redshifts as well \citep{tacconi20}.  Locally, there is a wide variation in the mass fraction of the neutral ISM that is found in the molecular gas phase with $R_\mathrm{mol} \equiv \Sigma_\mathrm{mol}/\Sigma_\mathrm{atom}$ ranging from $<10^{-1}$ to $>10^2$  \citep[][]{wongblitz02, blitz04,leroy13,schruba11, eibensteiner24}. However, nearly all of our insights into galaxy evolution focus on cases with $R_\mathrm{mol}>1$ where most of the neutral ISM is found in the molecular phase, all of which have bolstered the relatively simple view on star formation indicated above.  These regions are intrinsically bright in star formation and gas tracers making them comparatively easy to study.  

This focus emphasizes the actively star forming regions along massive galaxies in the ``star formation main sequence.''  These are, indeed, the environments where most of the star formation in the $z=0$ Universe are taking place.  However, it necessarily ignores both large areas in the outskirts of galaxies as well as populations of molecule-faint galaxies. Regions like our solar neighborhood are outside of the zone of regard for such studies, as are dwarf galaxies which can be intrinsically faint in their molecular gas tracers because of their low metallicities \citep{araa_xco, hunter_araa}, and the star forming discs of early-type galaxies, which can show a wide range of atomic gas fractions \citep[e.g.][]{Serra2012}.  All these systems are nonetheless actively star forming \citep{Roychowdhury2014} and present an essential contrast to the steady state star formation cycle that appears to typify the massive disks.  In such cases, the main regulators of the star formation process are (1) the amount of gas present but also (2) the conditions in the galactic environment that convert that gas into the cold neutral and molecular phases of the ISM \citep{saintonge16,saintoge22}.

The phase balance of the ISM is thought to depend on the local galactic environment, and principally on the gravitational potential that confines the gas to the galaxy.  Strongly confined gas has a higher volume density equilibrium, which favors cooling processes and thus efficient formation of the cold media \citep{fgh69,mckee77,draine11}.  There is good evidence for this general picture in the Milky Way through the gold-standard absorption line studies \citep[e.g.,][]{heiles03,Kanekar2011,Roy2013,murray18}, though these studies find significant fractions ($\sim 20\%$) of atomic gas mass in thermally unstable conditions, suggesting rapid cycling between ISM phases. Extragalactic 21-cm absorption line studies are more limited ~\citep[][]{dickey1993, dickey2000,Allison2021,dempsey2022} though the SKA precursors and pathfinders are paving the way for the SKAO to transform this view (see for example, the MeerKAT Absorption Line Survey, MALS, \citealt{Gupta2025}, or the GASKAP-HI survey, \citealt{chen2025}).  Attempts to decompose the atomic medium based on line width have faced both spatial and spectral resolution challenges \citep{stilp13, koch18}.


We are thus left with a picture that the atomic medium plays a central role in the star formation process, particularly in the outskirts of galaxies and in dwarfs.  However, our study of the physical processes that govern the role of atomic gas in star formation is limited to the Milky Way and a handful of the nearest galaxies. Extragalactic studies are particularly important because the outside perspective makes it dramatically easier to place the atomic gas observations in the broader galactic context.  In particular, they allow to directly compare the distribution of the different phases of the ISM and determine the spatial and kinematical relationship between the ISM, the stars formed from it, and the other components of a galaxy.


\section{Motivation}

In the past decade, a refined picture of galaxy evolution has emerged from high resolution observations ($\lesssim 100$~pc) of the molecular and ionised ISM across a large sample of galaxies, dramatically improving on the previous generation of studies at kpc scales. These differences in resolution match physically distinct scales: kpc-scales in galaxies can resolve the scale \textit{lengths} of galaxy disks whereas 100-pc scales resolve the typical scale \textit{heights} of galaxies. Resolving a scale length indicates how a galaxy is organized radially and illuminates broad evolutionary processes that describe how matter moves through disks and how galaxies build up their mass, but this view treats matter in terms of vertically integrated quantities (surface densities). It is necessary to resolve the scale heights of galaxies, reach the scales where the ISM can be treated in terms of its three-dimensional structure, thus making meaningful measurements of the intrinsic volumetric quantities that underpin theoretical models (volume densities). Further, the tight ``star formation law'' between gas and star formation \citep[e.g.,][]{kennicutt_evans} breaks down on these scales with star formation appearing spatially distinct from the fueling molecular gas \citep{schruba10,onodera10, kruijssen14}. Thus, 100-pc scales represent a key threshold to reach for transformational observations of the star forming ISM across a representative sample of galaxies. 

High resolution observations enabled the progress outlined in the introduction \citep{schinnerer24} but there remains a pressing need for similar observations of the atomic gas in galaxies.  These observations are essential in the atomic-dominated parts of the ISM. The long wavelength of the \textsc{H~i} 21-cm line, the best tracer of the atomic ISM, means that radio interferometers are needed for high resolution (100 pc) observations.  To access a broad range of galaxy types and environmental conditions, we must be able to achieve this resolution out to a distance of 20 Mpc.  This volume includes 1400 star forming galaxies with $M_\star >10^8~M_\odot$, and a similar number of systems in the various stages of quenching their star formation. It also spans a range of environments in which those galaxies reside. 

Before the SKAO pathfinders and precursors, kpc resolution \textsc{H~i} observations could reach galaxies in the Virgo cluster \citep[$D= 16$~Mpc, see for example the VLA Imaging of Virgo spirals in Atomic gas, VIVA][]{viva_survey}, linking the detailed evolution of the star forming ISM to how the broader galaxy cluster environment reshapes star formation, and allowing access to the most massive galaxies in our local universe.  \textsc{H~i} observations of the nearby clusters Coma and Virgo provided one of the first evidences ram-pressure from the environment stripping the gaseous disk of galaxies~\cite[][]{Gavazzi1989,Kenney2004}. Recently, in these extended tails, thanks to the new observing facilities, it has been possible to detect and study new episodes of star formation~\citep{Poggianti2019}.  Recently, the SKA-Mid precursor MeerKAT has allowed to expand these studies to higher distances and lower density environments such as the Fornax cluster~\citep[$D=20$~Mpc,][]{Serra2023,Serra2024}.  With resolved studies of \textsc{H~i} in nearby galaxies and their environment as a science driver, we would need to be able to observe the \textsc{H~i} 21-cm line with $1''$ resolution (80\,pc at the distance to Virgo; 100 pc in Fornax). 

\begin{figure}[ht!]
    \centering
    \includegraphics[width=0.7\linewidth]{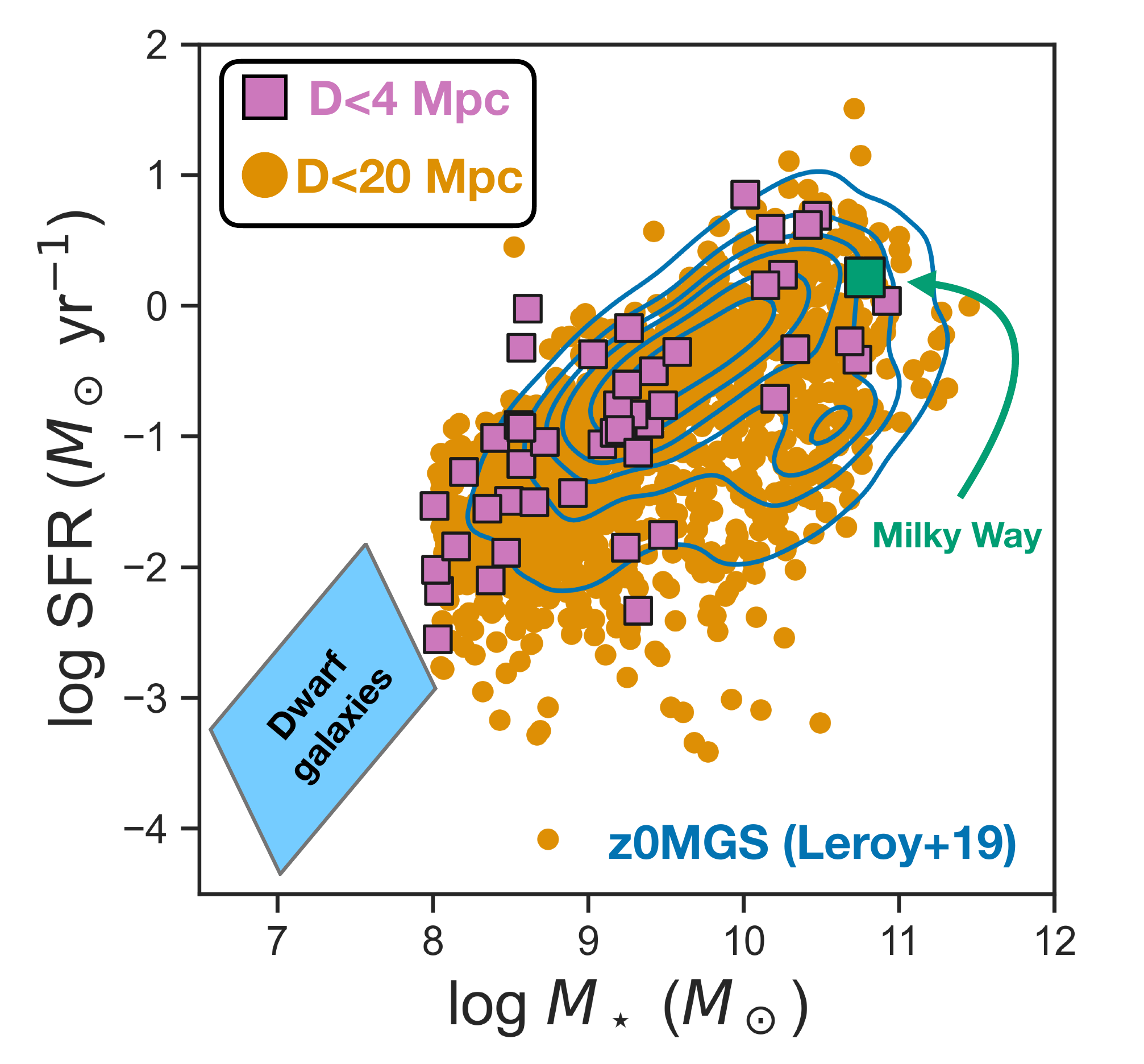}
    \caption{Massive galaxies from the z0MGS sample \citep[][complete to M$_\star > 10^8$~M$_\odot$ within $d=50$~Mpc]{z0mgs} showing the available populations of galaxies where current radio interferometers can reach 100~pc resolution ($D<4$~Mpc) and the vastly larger and representative $z=0$ population that $1^{''}$ {\sc HI} mapping with SKA-Mid AA4 can reach ($D<20$~Mpc), including a representative dwarf galaxy populations (M$_\star < 10^8$~M$_\odot$). By reaching $1^{''}$ scales with {\sc HI} mapping, SKA-Mid AA4 enables resolved {\sc HI} studies on 100~pc scales in Virgo, the nearest cluster environment.}
    \label{fig:sfms}
\end{figure}

Figure \ref{fig:sfms} compares the samples that can be obtained by the current generation of facilities, which obtain high quality observations\footnote{We define ``high quality'' as meaning an rms column density sensitivity $N_\mathrm{HI}<10^{20}~\mathrm{cm}$ in 10~$\mathrm{km~s^{-1}}$ line width.}
out to $D\sim 4$~Mpc, to the samples that could be observed with SKA-Mid.  The survey volume and numbers of targets available grow with $D^3$.  The nearby galaxies lack a broad sampling of the star formation patterns seen across galaxies as a whole.  Notably absent are the ``quenched'' or ``retired'' populations of galaxies.  The aforementioned lack of any galaxy cluster environments also misses one of the key drivers of galaxy evolution \citep{cortese21}.  This is especially important because the cluster environment most dramatically affects the atomic neutral medium~\citep[see for example the Virgo, Fornax and Coma cluster, e.g., ][respectively]{viva_survey,Serra2023,Molnar2022} with less influence on the molecular gas \citep{fumagalli09,watts23}. While less obvious, galaxies in lower density groups also have their atomic medium affected by the environment \cite{brown17}.

Unfortunately, $\sim$ 100-pc resolution alone is insufficient.  The key metric is surface brightness sensitivity on these small angular scales, which requires substantial sensitivity on baselines of $\sim 40$~km. The 21-cm line emission is a thermal process, so it is limited to low received power, especially from the cold phases of the ISM so tightly linked to star formation.  Ambitiously, we propose targeting a column density sensitivity of $10^{20}~\mathrm{cm}^{-2}$, while resolving a 10 km/s line width with 100-pc resolution.  This is readily exceeded in the Local Group with current facilities and will be expanded to nearby targets $D<4$~Mpc with AA$^\star$. However, using these facilities to extend to more distant targets is punishing.  Using a simple scaling, the observing time required for a given facility to achieve a specified surface density sensitivity on a fixed \textit{physical} scale scales like $D^4$ where $D$ is the distance to the target.  This harsh scaling stems from the conversion from an approximately constant flux density sensitivity ($\sigma_S$) to a brightness temperature sensitivity ($\sigma_T$) that depends on beam solid angle ($\Omega$): $\sigma_T \propto \sigma_S/\Omega$.  Since the required solid angle of the beam to achieve a required resolution scales like $\Omega \propto D^{-2}$ and the noise in an image $\sigma_S \propto t^{-1/2}$, a constant $\sigma_T$ requires $t\propto D^4$.  Increasing surveys to a large number of galaxies requires more sensitive facilities, which practically requires the additional collecting area that AA4 provides.

\begin{figure}
    \centering
    \includegraphics[width=0.7\linewidth]{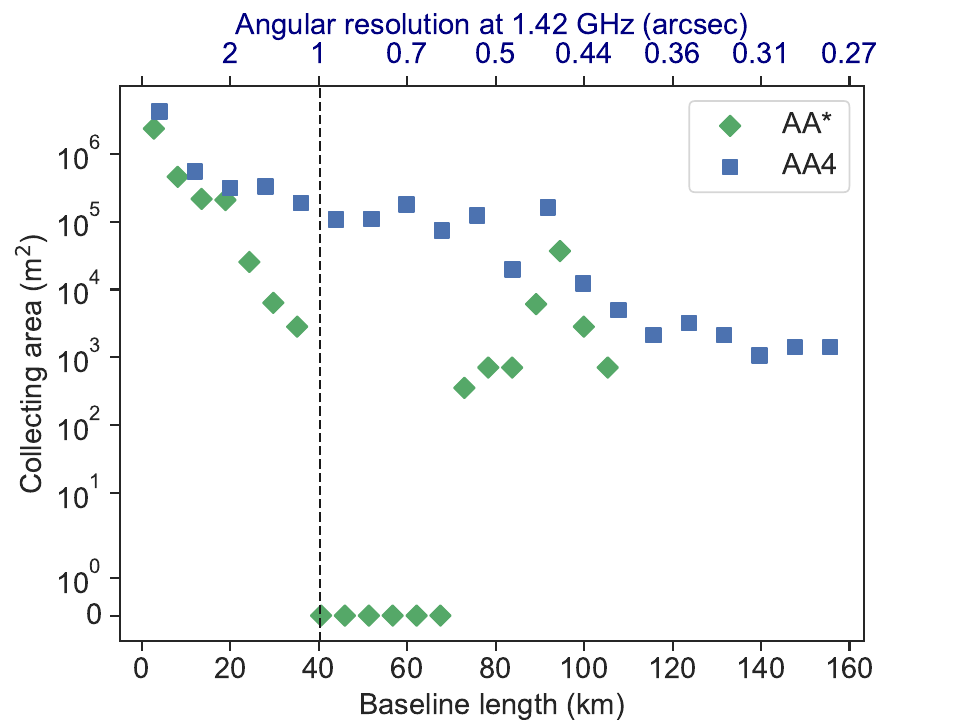}
    \caption{Comparison of the collecting area by baseline length between AA$^\star$ and AA4 for SKA-Mid. The top shows the equivalent angular resolution at 1.42~GHz for the 21-cm {\sc HI} line. Achieving $1^{''}$ {\sc HI} mapping, equivalent to 100~pc in the Virgo Cluster (20~Mpc), is uniquely suited to reaching the full AA4 configuration for SKA-MID.}
    \label{fig:aastar_vs_aa4_collectingarea}
\end{figure}

Figure \ref{fig:aastar_vs_aa4_collectingarea} illustrates this for SKA-Mid under AA$^\star$ and AA4 configurations.  While AA4 of SKA-Mid will improve on collecting area at all baselines, it will most dramatically increase collecting area from 20 to 70 km baselines, which makes sensitive 21-cm observations at $1''$ possible (obtaining a 5–10$\times$ resolution gain at comparable column densities of VIVA). The AA$^\star$ observations will provide excellent resolved-gas studies of individual galaxies out to 10 Mpc, which will improve our statistical treatment of the relation between atomic gas and star formation.  However, transformational studies will require observing targets out to $D=20$ Mpc. In aspiring for this targets, we can access an order of magnitude larger galaxy population by number and study galaxies inside galaxy clusters including spanning the Virgo cluster ($D=16$~Mpc) and the near side of Fornax ($D=20$~Mpc). A complete picture of galaxy evolution necessary requires studing the environments of galaxies from gas rich groups, to clusters, to voids.

\begin{figure}
    \centering
    \includegraphics[width=1.0\linewidth]{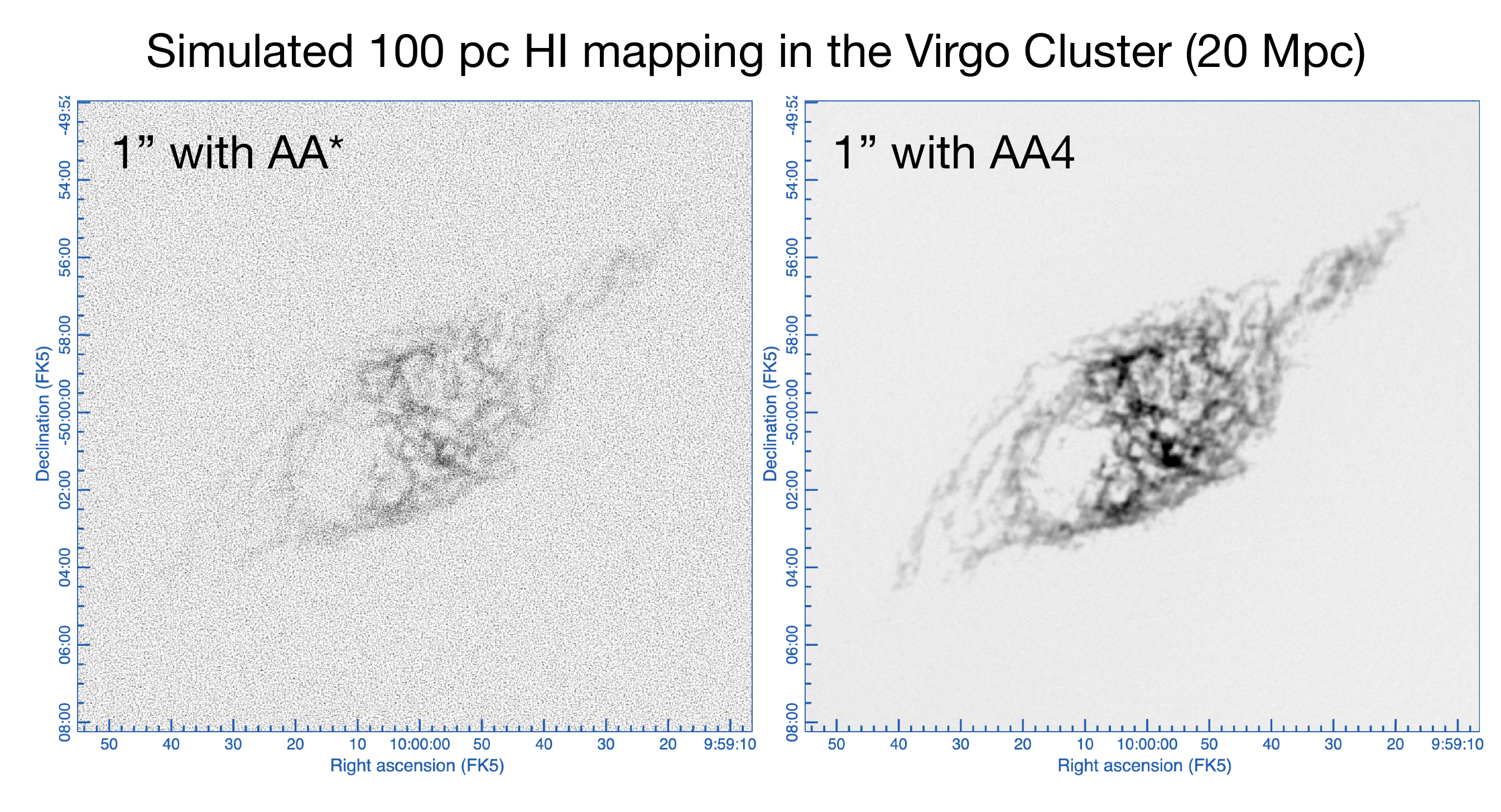}
    \caption{Simulated $1^{''}$ SKA-MID 21-cm {\sc HI} observations of NGC~6822 placed at the distance to Virgo (20~Mpc) using high-resolution VLA mapping from LGLBS \citep[][Pingel et al. in preparation]{lglbs_survey}. Both simulations use noise levels derived from a channel width of 10~km/s and 100~hr on source with AA$^\star$ (left) and AA4 (right) configurations with varying weighting schemes to reach the target $1^{''}$ beam size. The AA4 simulated image matches the highest-resolution LGLBS column density sensitivity of $1\times10^{20}$~cm$^{-2}$ that represents the best-possible sensitivity of the VLA with a $4.5^{''}$ beam.
    The noise level in AA$^\star$ is $13\times$ higher than in AA4 with the same observing time, making it unfeasible to improve the above simulated image due to the lack of long baseline coverage.
    \textbf{For resolved {\sc HI} surveys across nearby galaxies, AA4's long baseline coverage is essential for enabling this science.}}
    \label{fig:ngc6822_in_virgo}
\end{figure}

Figure \ref{fig:ngc6822_in_virgo} vividly illustrates the need for AA4 in making galaxy cluster observations of well-resolved ($1''$) atomic gas maps ($D=16$ to 20 Mpc). These maps are completely out of reach with current facilities because of their low surface brightness sensitivity or coarse resolution \citep[e.g.,][which reaches good sensitivity but at $30''$ resolution]{Boselli2023}

Even with 100 h integrations on target, the AA$^\star$ observation requires aggressive weighting on long baselines without substantial collecting area yielding poor quality maps.  AA4 will make substantially better maps of galaxies in these essential cluster environments. This figure represents an extreme case: 100-pc resolution for a galaxy at 20 Mpc, which would be able to observe the Fornax cluster along with Virgo thus expanding detailed HI studies to lower density environments.  The challenges can be alleviated by observing nearer targets.  As an extreme illustration, 100 pc-scale observations for a galaxy at 4 Mpc would only require 10 minutes of observing time with the respective arrays. Indeed, nearby targets are already being observed with precursor facilities~\citep[e.g. MHONGOOSE, LGLBS,][]{mhongoose_survey,lglbs_survey}.


\section{Precursor Studies}

Observations of the nearest targets illustrate the scientific insights to be gained from high resolution observations. Milky Way and Magellanic Cloud studies reveal the complex spatial structure from AU to kpc scales \citep{stanimirovic_araa} and spectral structure ranging from the thermal line width (1 km/s) of the cold neutral medium up to hundreds of km/s associated with galactic rotation and inflow/outflow. These details are lost in studies of more distant targets. The previous generation of extragalactic \textsc{H~i} surveys with $\sim$kpc-scale resolution, like 
the WHISP~\citep{Swaters2002} THINGS \citep{things_survey}, FIGGS \citep{Begum2008}, LITTLE-THINGS \citep{little_things} HALOGAS~\citep{Heald2011}, VIVA \citep{viva_survey} and LV\textsc{H~i}S~\citep{Koribalski2018}, were transformational in illustrating the distribution and kinematics of atomic gas in galaxies.  However, the quality of their data could only sustain simple views of the ISM: smooth ($15''$ resolution) \textsc{H~i} distributions and 1 or 2 component Gaussian spectral models, representing at most, the cool and warm neutral gas. While molecular gas studies have realized an order-of-magnitude jump in their spatial resolution (and sensitivity), analyses that combine the atomic and molecular gas frequently resort to asserting the smoothness of the atomic gas on kpc-scales \citep[e.g.,][]{sun_2020} or assuming the atomic gas has a uniform surface density of $\Sigma_\mathrm{atom}\sim 10~M_\odot~\mathrm{pc}^{-2}$ ($N_{\rm HI}\sim 10^{21}$ cm$^{-2}$) over the studied region. 

In the past decade, there have been continued efforts to improve our samples of nearby galaxies, but these remain limited by the capacity of the observatories at work. Combining the VLA B,C, and D configurations achieves good surface brightness sensitivity but only to scales of $\sim 8''$ for 21-cm observations. The A-configuration is sparse are requires significant observing time to improve the angular resolution below $6''$. 
Similarly, good surface brightness sensitivity observations can be done with GMRT down to $\sim 5''$. The SKA-precursor MeerKAT has excellent $uv$ sampling at small baselines, spanning the same $uv$ distances as VLA B+C+D configuration imaging but with more sensitivity than the VLA or GMRT on these scales. Even so, MeerKAT is also limited to $\sim 6''$ scales for good sensitivity of observations. Hence, there have not been significant gains in the ability to make $1''$-scale maps of atomic gas.  We have, however, dramatically expanded the sample of galaxies with lower resolution \citep[e.g.,][]{laudage24,eibensteiner24,degasperin2025}.  Further, major investments of time to observe galaxies have expanded our awareness of the low-column density gas in the outskirts of galaxies, notably the MHONGOOSE survey on targetted nearby SF galaxies ($<20$~Mpc) \citep{mhongoose_survey} and the MeerKAT Fornax Survey and VICTORIA on nearby clusters~\citep[][]{Serra2023,Boselli2023,degasperin2025}. 

\subsection{What High Resolution Observations Reveal}

\begin{figure}
    \centering
    \includegraphics[width=0.8\linewidth]{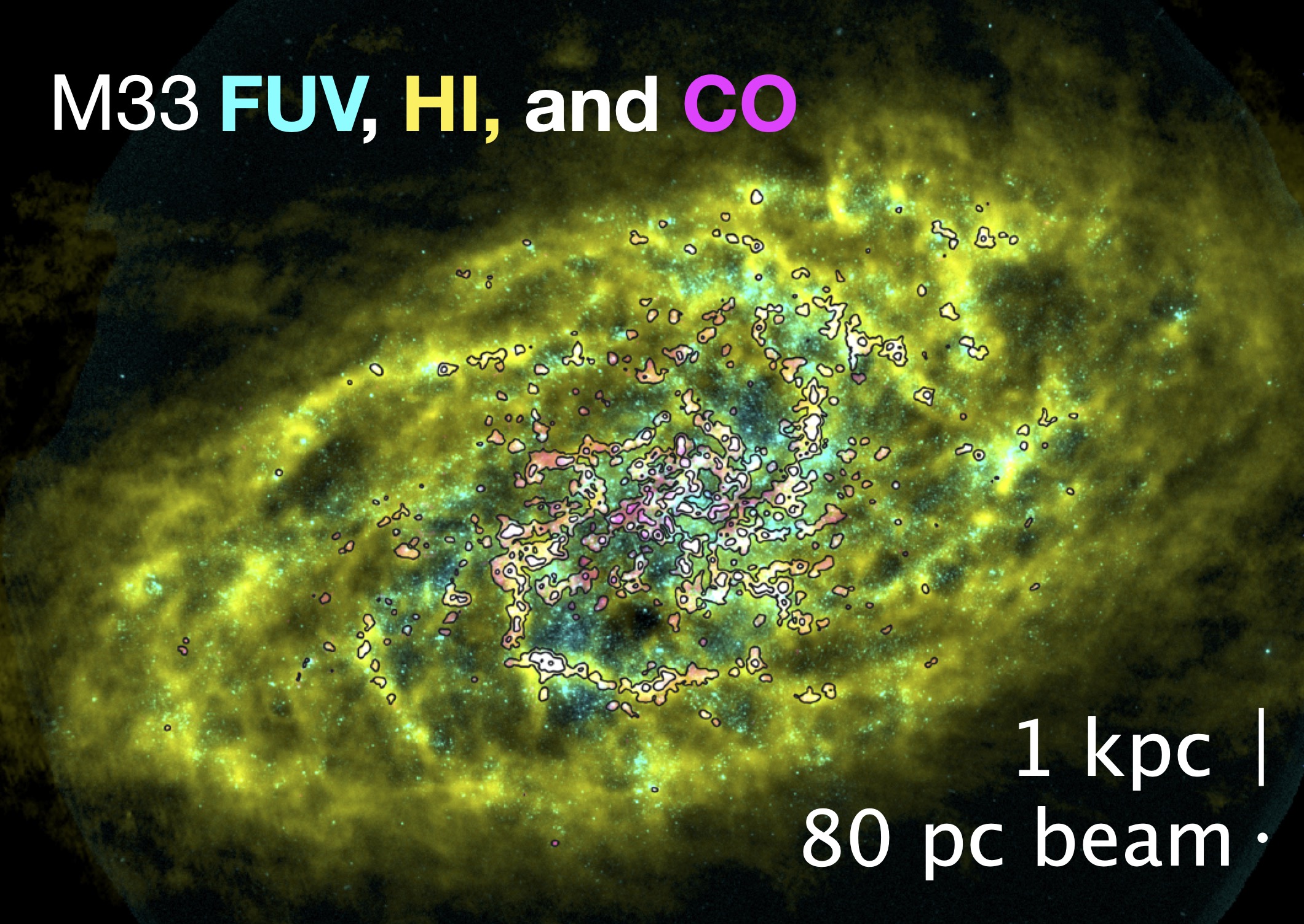}
    \caption{The Local Group galaxy M33 shows the integrate intensity for \textsc{H~i} 21-cm emission from the LGLBS \citep[yellow;][]{lglbs_survey}, far ultraviolet emission from GALEX \citep[cyan;][]{galex_atlas} and CO(2-1) emission \citep[magenta and contours;][]{gratier10}. The atomic and molecular gas data have a projected beam size of 80 pc and the 1 kpc scale bar indicates the typical resolution scale for data on nearby star forming galaxies. }
    \label{fig:m33-lglbs}
\end{figure}

We can, however, still obtain sensitive $\lesssim$ 100-pc scale observations in nearer targets beyond the Milky Way and Magellanic clouds. The next obvious set of targets are more distant Local Group targets, which have been extensively mapped in \textsc{H~i} \citep{wilcots98,thilker04,koch18, braun09} at this resolution.  
Beyond that up to a distance of 3.5 Mpc, the \textsc{H~i} in star-forming dwarf galaxies from the FIGGS survey have been mapped at 200~pc scales for the purpose of studying the relationship between atomic gas and star formation \citep[][see also Section~\ref{ssec:kslaw}]{Roychowdhury2009}.
More recently, the VLA Local Group L-Band Survey \citep[LGLBS; ][]{lglbs_survey}, which used VLA ABCD imaging to map the targets in atomic gas and the radio continuum.  While imaging data of the 21-cm line at $4.5''\approx 20~\mathrm{pc}$ is forthcoming, the survey has released $\sim 100$ pc scale maps.

Figure \ref{fig:m33-lglbs} shows the relationship between \textsc{H~i}, molecular gas emission and star formation as traced by the far ultraviolet emission in M33, which provides a reasonable model for the relationship between these three tracers in a star forming disk galaxy. The most immediate result from this image is the co-location of the atomic and molecular gas, as expected since the molecular gas forms from the atomic medium, rapidly forms stars, and is disrupted by stellar feedback.  A close correlation between the atomic and molecular phases is expected for short lived molecular clouds.

In the central several kpc of the galaxy, the star formation is clearly associated with the molecular gas emission. There are clear ``bubbles'' visible in the atomic gas that are clearly connected to FUV emission and are the signs of stellar feedback. These bubbles are best traced using atomic gas since it fills a larger volume of galaxies than the molecular ISM and, with a lower density, is more responsive to the stellar feedback. Spectrally resolved atomic gas measurements also provide important kinematic tracers of super-novae feedback, which can directly measure the momentum and energy in expanding shells \citep{bagetakos11}.

In the center of the galaxy, the atomic gas emission becomes weak compared to the molecular gas emission showing a transition to a molecule- and feedback-dominated ISM. At larger radius, the atomic gas emission becomes stronger, both in an absolute sense and relative to the molecular gas. Here, there are several star forming complexes visible in the FUV that are most closely associated with the atomic gas.  These stars almost certainly still formed in molecular clouds with efficiency comparable to the molecule-rich parts of galaxies \citep{schruba11}, but those clouds are either faint because of reduced CO emissivity at low metallicity or they have already been disrupted by feedback. Either way, the pervasive atomic gas remains the tracer that highlights the ISM from which the stars are forming.

All of these immediate results that we glean simply by looking at these resolved maps become washed out at kpc-scale resolution.  The broad correlations will persist at coarse scales \citep[e.g.,][]{boquien15}, but the ability to discern structure, establish the relationship between atomic and molecular gas, and connect the response of the ISM to stellar feedback is lost at coarser resolution.

\subsection{Star Formation and Atomic Gas}
\label{ssec:kslaw}

\begin{figure}
\centering
\includegraphics[width=5in]{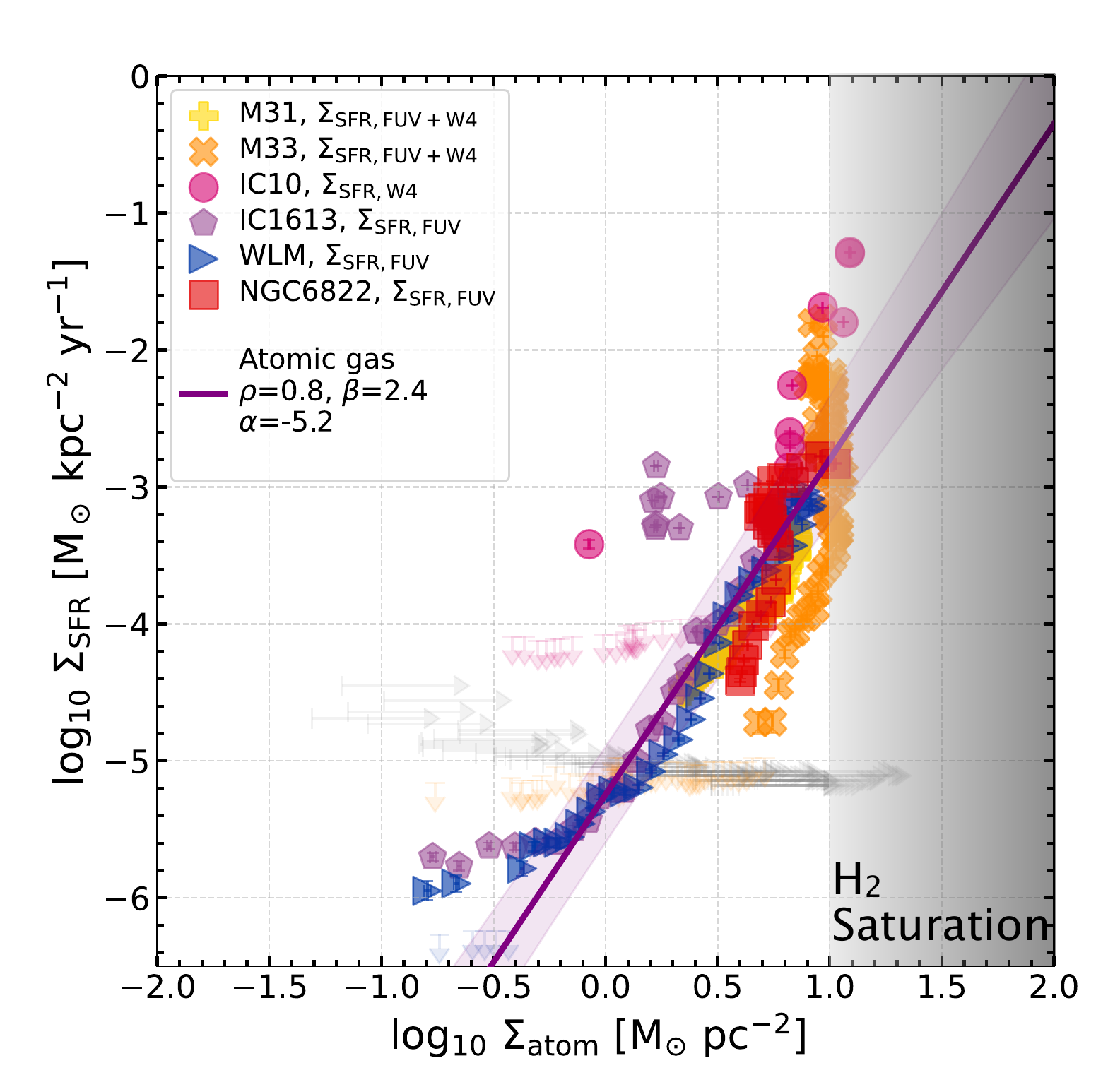}
\caption{Resolved star formation law on 120 pc in scales in Local Group Galaxies \citep{eibensteiner26}.  This figure shows the star formation rate as a function of the atomic phase surface density (including helium).  Star formation rates are derived from far ultraviolet and WISE 22 $\mu$m image following \citet{z0mgs}. Faint points indicate limits for the corresponding data. Star formation rates are correlated with atomic gas surface up to $\Sigma_\mathrm{atom}\approx 10~M_\odot~\mathrm{pc}^{-2}$, above which the neutral ISM becomes dominated by the molecular phase of the ISM.} 
\label{fig:ks-lglbs}
\end{figure}

\begin{figure}
\centering
\includegraphics[width=5in]{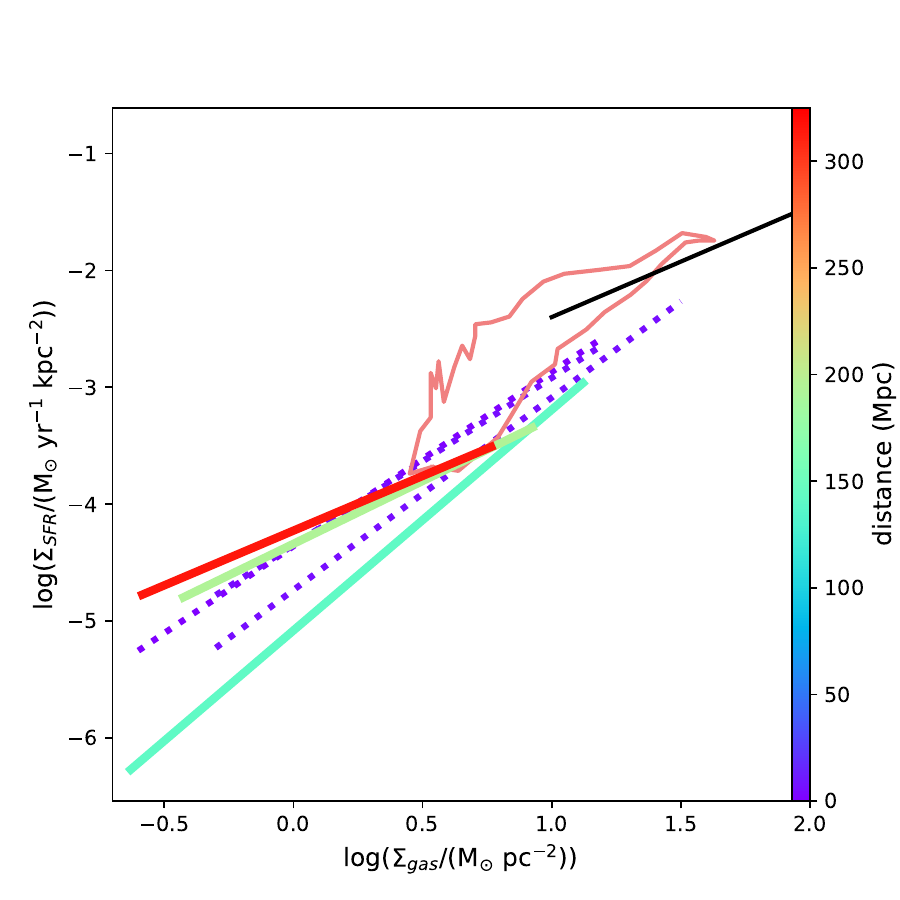}
\caption{The resolved atomic gas Kennicutt-Schmidt (K--S) relations as measured using the mean SFR surface densities in \textsc{H~i} column density bins, for \textsc{H~i}-domiated ISMs of dwarf galaxies and outer disks of massive spirals colour coded by median distance to the samples, contrasted with the molecular K--S relation in black \citep{leroy13}, and the total gas K--S relation within the optical disks of THINGS spiral galaxies outlined by the brown contour \citep{Bigiel2008}. The dashed lines are from \citet{Roychowdhury2015} for FIGGS dwarf galaxies and outer disks of THINGS spiral galaxies at 400 pc resolution (virtually indistinguishable from each other) and for THINGS spirals at 1 kpc resolution (lower efficiency). The bold lines are from the recent study using MeerKAT (Roychowdhury et al. in preparation), where the increasing distances also correspond to increasing median resolutions of 7, 10 and 17 kpc respectively.} 
\label{fig:kssr}
\end{figure}

We can directly quantify the relationship between the atomic gas and star formation, as demonstrated in recent works, including \citet{Wagh2024}. Figure \ref{fig:ks-lglbs} shows the resolved star formation law for atomic gas in the Local Group, relating the surface density of atomic gas to the local star formation rate.  Just as is true with molecular gas at high surface densities, there is a clear power-law relationship between the amount of atomic gas present in a system and the star formation rate, but this trend only holds at low surface densities of gas: $\Sigma_\mathrm{gas}\lesssim 10~M_\odot~\mathrm{pc}^{-2}$. Above this threshold, the ISM becomes saturated with molecular gas and the better studied molecular gas star formation law emerges \citep{leroy13, kennicutt_evans}.  The relationship also appears steeper than the molecular gas star formation law: here $\Sigma_\mathrm{SFR}\propto \Sigma_\mathrm{atom}^{2.4}$ while we usually expect $\Sigma_\mathrm{SFR}\propto \Sigma_\mathrm{mol}^{1.0}$.  We also see significant galaxy-to-galaxy offsets and the transition to a molecule-dominated ISM occurs at slightly different surface densities.
A steeper star formation relation has also been seen in few-hundred pc to kpc scale resolved studies of dwarf galaxies and outer regions of massive spirals within the Local Volume \citep{Roychowdhury2009,Roychowdhury2011,Bigiel2010}.

At the same time, the Kennicutt-Schmidt relation as defined by the optical disk-averaged values of the surface densities of \textit{total} gas and SFR appear to form a universal relation across (non-starbursting) galaxy types, even when dwarf galaxies are included for which atomic gas dominates the total gas mass \citep{delosreyes19}.
There seems to be a mismatch between the empirical determinations of the star formation relation involving atomic gas for galaxy-averaged and resolved studies.
When measuring the resolved relation in \textsc{H~i}-dominated ISM with low SFRs, measuring the SFR itself becomes an important caveat \citep{daSilva2012}.
\citet{Roychowdhury2015} considered this and used the mean value of the SFR in \textsc{H~i} column density bins from sub-kpc and kpc scale measurements of the same in dwarf galaxies and outer disks of massive spirals within the Local Volume.
They not only found that the Kennicutt-Schmidt relation was identical for the two different samples at the same spatial scale, but less steep than previously measured and matching the expectations from models where the physical conditions in the ISM are set by stellar and supernova feedback (Figure \ref{fig:kssr}).

MeerKAT, though not improving the spatial resolution when compared to the previous generation of radio interferometers, is enabling \textsc{H~i} observations at much further distances because of its improved sensitivity.
A study similar to the above done using the preliminary data from MIGHTEE-HI a large MeerKAT survey~\citep{Maddox2021} has looked the variation with redshift of the resolved Kennicutt-Schmidt relation in the outer disks of massive spirals up to $z \sim 0.1$, though over scales ranging from few kpc to tens of kpc (Roychowdhury et al. in preparation).
Whereas \citet{Roychowdhury2015} clearly demonstrated that lowering the spatial resolution for the same sample makes the star formation law artificially less efficient,
intriguingly the recent study finds that even though the spatial resolution becomes coarser with redshift, the star formation relation becomes more efficient (Figure \ref{fig:kssr}), hinting at the possibility of more efficient star formation in the outer disks of spirals with increasing redshift.
SKA will start confirming such trends using methods similar to above but with much better resolutions and to higher redshifts.
And for the Local Volume galaxies which can be resolved down to their scale heights, SKA will be able to directly quantify underlying the volumetric Kennicutt-Schmidt relation in the \textsc{H~i}-dominated ISM 
\citep{Bacchini2020, Wagh2024}.

Recent studies by \citet{Wagh2024} have contributed significantly to advancing this picture by investigating the connection between atomic gas and star formation in molecular hydrogen-rich galaxies (MOHEGs) at redshifts, $z <0.2$ \citep{Ogle2010}. Their study shows that although molecular gas remains the primary correlate of high star formation rates, atomic gas is detected in roughly $70\%$ of the sample and plays a substantial role in regulating star formation efficiency. \citet{Wagh2024} suggest that this steeper relationship indicates a distinct mode of star formation in atomic-dominated regions that is not fully explained by classical star formation models based on molecular gas. These results highlight the need for high resolution H\,{\sc i} observations that the SKA will deliver.  

These observations illustrate that star formation in the atomic-dominated parts of the ISM presents an intriguing contrast to the molecule-rich disk mode that has been heavily studied to date.  These atomic-dominated regions are found in outer regions of massive star forming galaxies or throughout dwarf galaxies. For nearby galaxies, these are also regions are also those for which we will gain the most insight into the stellar populations from the multiwavelength observations making wide area surveys from facilities like \textit{Euclid, Roman} and Rubin. In the future, wide field ($10'$) integral field spectrographs will allow us to correlate the kinematics of the stars and of the ionised ISM with the SKA-Mid AA4 observations of the \textsc{H~i}.

\subsection{Key Science Questions}

High resolution observations of the atomic medium in a range of galaxies will answer key unknowns for which the SKA-Mid, especially AA4, will be required.

\paragraph{How does the molecular medium form?} Theoretical models and observations of the ISM support the idea that molecular gas forms out of the cold neutral atomic gas
\citep{mckee07} but the details of this process remain unclear, especially as they pertain to star formation.  One major distinction between the diffuse and the star forming ISM is the role of self-gravitation.  Viewed through the lens of the Local Group, the molecular ISM is also the phase of the ISM where self-gravity is a significant factor.  In more molecule rich regions like galaxy centers or more massive targets, the molecular medium is not necessarily synonymous with self-gravitation \citep{pety13}. 

While theoretical models understand the balance between the atomic and molecular gas in terms of photodissociation models \citep{kmt09}, but a complete picture requires predictions in how the conditions in the atomic gas lead to self-gravitating structures \citep{krumholz13}. To understand the atomic-molecular transition, we need to study atomic gas at high spatial resolution to understand the interface of cold neutral medium (CNM) formation and the changing dynamical properties of the ISM.  This is clearly a function of environment (e.g., galactocentric radius as in Figure \ref{fig:m33-lglbs}). How does galactic environment and the conditions in the atomic ISM create the CNM and what parts of the CNM go on to form molecular clouds? How efficient is the conversion of the CNM into star forming molecular gas? What are the properties of those molecular clouds and how are they related back to the atomic gas?

We also seek deeper insights on how galactic dynamical features shape star formation.  For example, spiral arms, bars, galaxy cluster environments, and galactic interactions all reshape a galaxy's ISM and star formation processes.  While the influence of these effects on molecular gas is studied on 100-pc scales \citep[e.g.,][]{pessa21}, their influence of on the atomic gas is poorly known since we are not able to resolve the atomic gas structure on similar scales.  How do spiral arms and bars change the properties and conditions of the atomic ISM?  How do these changes compare to what is seen in the molecular gas and star formation? How exactly do galactic interactions and galaxy cluster environments quench galaxies through their effect on the atomic ISM?

\paragraph{How efficiently does feedback connect to the ISM?} Stellar winds, radiation, and supernova shocks all inject energy and momentum back into a galaxy's ISM, collectively resulting in stellar feedback on a galaxy.  The efficiencies, timescales, and global influence of feedback remain the single largest unknown in understanding star formation and the evolution of isolated galaxies \citep{grudic2019}.  By identifying feedback structures in the atomic gas, we will be able to measure the kinematics, swept up mass, and sizes of expanding feedback shells and connect these gas structures back to their stellar populations that drove them.  This will provide the cleanest measurement of how much energy and momentum are deposited back into the galaxy.  This requires a large sample of feedback shells across a range of environments to understand how this coupling efficiency relates to the local galactic environment, the original ISM structure, and the mass, age, and metallicity of the driving stellar population. Since many SF galaxies in the nearby universe also show an active galactic nucleus (often Seyfert galaxies with also radio jets of low-radio power)~\citep[see, for example,][]{Mingozzi2019}. AGN feedback phenomena will also be investigated and it will be possible to compare their impact on the evolution of their host with respect to SF feedback (see Maccagni et al. in this book, for further details).


\section{Towards a statistical sample of nearby galaxies}

There are clear scientific opportunities for using the SKA to unravel the relationship between the atomic gas in galaxies and star formation. While AA$^\star$ will continue the evolution of our understanding, fully answering these key questions will require AA4.

\begin{figure}
    \centering
    \includegraphics[width=1.0\linewidth]{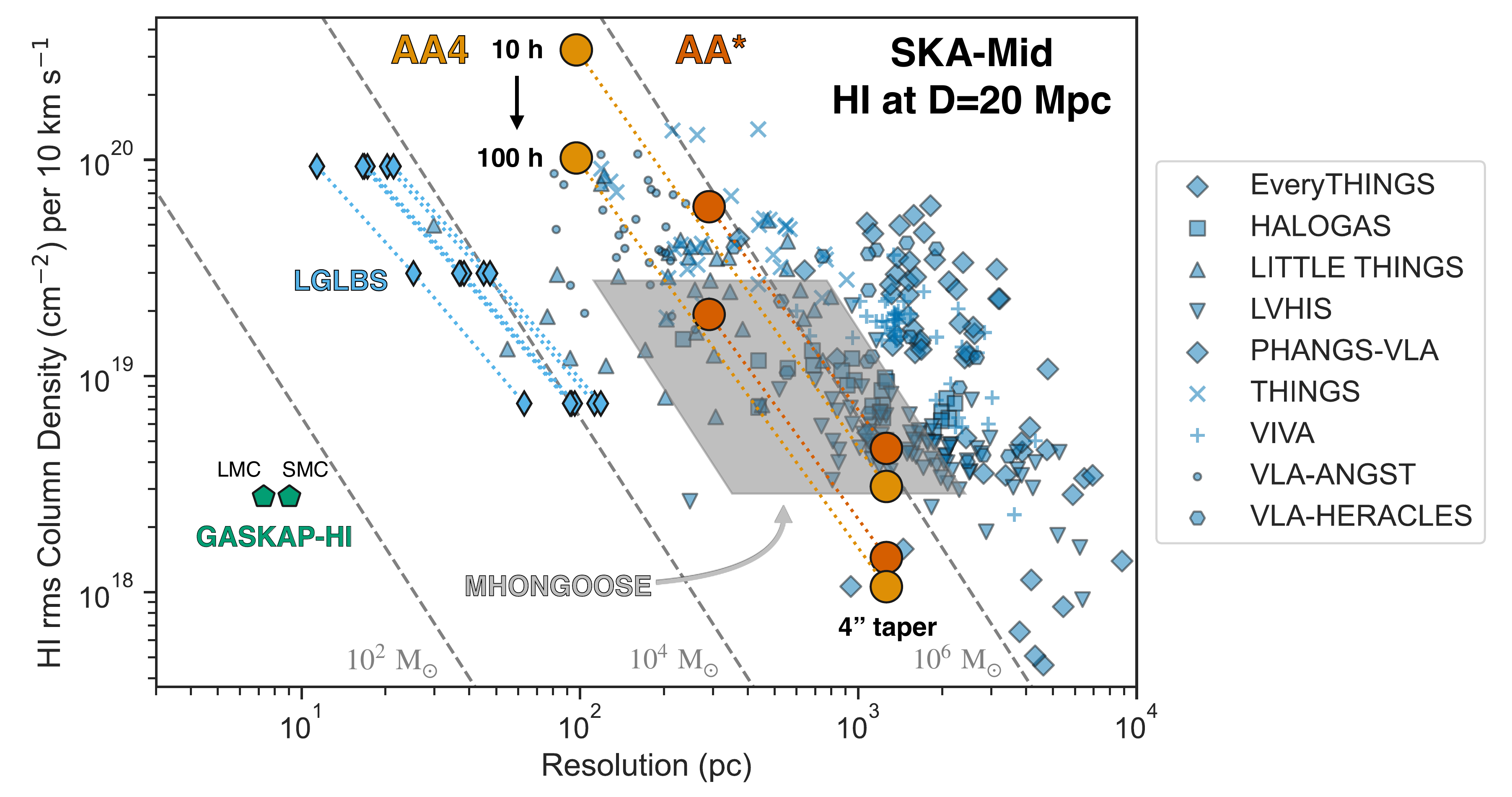}
    \caption{{\sc HI} column density sensitivity of current surveys compared to SKA-Mid predictions for a galaxy at 20~Mpc in the Virgo cluster~\citep[adapted from][]{Maccagni2024,lglbs_survey} The predicted SKA-Mid values use show AA4 and AA$^\star$ with integration times of 10 and 100 h. The connected lines show the trade-off of resolution and sensitivity between Briggs weights (robust$=1$) for high resolution and outer uv-tapering with a $4''$ Gaussian taper for high sensitivity.
    Current surveys are shown for galaxies at a range of distances, from within the Local Group (GASKAP-HI and LGLBS) to $D=25$~Mpc with the MHONGOOSE sample~\citep{mhongoose_survey} and various surveys indicated in the legend (
    EveryTHINGS, I. Chiang et al. in preparation;  
    HALOGAS, \citealt{Heald2011}; 
LITTLE THINGS, \citealt{little_things}; 
LVHIS, \citealt{Koribalski2018};  
PHANGS-VLA, presented in \citealt{sun_2023};
THINGS, \citealt{things_survey}; 
VIVA, \citealt{viva_survey}; 
VLA-ANGST, \citealt{Ott2012}; 
VLA-HERACLES, \citealt{schruba11}; 
MHONGOOSE, \citealt{mhongoose_survey}), and LMC and SMC (e.g., GASKAP-HI, \citealt{Pingel2022}), which are $>10\times$ nearer than any of the LGLBS targets.
See also \citet{Maccagni2024}. Relative to VICTORIA, the current state-of-the-art MeerKAT survey of the Virgo Cluster \citep{degasperin2025}, SKA-Mid offers a $5\mbox{--}10\times$ linear resolution gain at comparable column density sensitivity.}
    \label{fig:ska_vs_current}
\end{figure}

Figure \ref{fig:ska_vs_current} illustrates how observations of the SKA toward nearby galaxies maps effectively into the relevant physical parameter spaces needed for highly resolved studies of the ISM adapted from ~\citet{lglbs_survey}. For a complete census of {\sc HI} surveys in nearby galaxies see ~\citet{Maccagni2024}. The figure illustrates both AA$^\star$ and AA4 expected sensitivity for 10 h and 100 h observation times toward single targets under different weighting schemes for targets at a distance of 20 Mpc.  The comparator surveys shown as points illustrate how several different surveys relate to these efforts.  Baseline observations with the SKA will be comparable to previous generation studies of modestly closer targets.

The reality is that sensitive observations of atomic gas at 100 pc resolution at the far edge of the Virgo cluster or at the distance of Fornax requires a substantial investment of array time, even for AA4.  A large investment of 100 h would need to be reserved for essential targets that illustrate unique parts of parameter space.  This represents an outer limit that spans the necessary targets to get a full perspective on galaxies, both in their star formation properties and in their cluster environment.  Observing closer targets requires dramatically less time to reach reasonable column densities at high resolution.

An SKA-mid survey that is designed to answer the motivating questions about the relation of atomic gas to star formation would consist of several tiers.  With AA4, the core sample would likely consist of every visible massive galaxy within 10 Mpc, which could be surveyed at 100 pc resolution to good sensitivity (rms column density of $10^{19.3}~\mathrm{cm}^{-2}$) in modest time ($\sim 6$ hours). This core survey would consist of at least 100 targets (for $M_\star>10^8~M_\odot$) and could include several more factors to sample lower mass dwarf galaxy targets. Nearer targets will achieve even higher physical resolution, but extending across the Virgo and Fornax clusters requires either heavy investments of time or compromising on survey metrics. Both cluster and isolated galaxies will be required for a complete picture, identifying a few targets for detailed observation while spanning a large number of targets to gain a statistical perspective.  Many of these targets already have substantial investment from complementary multiwavelength facilities: ALMA, HST, JWST, VLT/MUSE, etc.~\citep{schinnerer24}. Only the atomic gas -- the dominant mass phase of the ISM -- is missing for a near-complete complete picture of the matter in these galaxies.

Precursor data for these surveys exist now.  With the closest targets ($D<4$~Mpc), current facilities produce maps with the requisite resolution.  With the Milky Way and Magellanic Cloud studies untangling physics at the smallest scales, the Local Group providing an efficient bridge to different galaxy types and this larger ``Local Volume'' sample, we can carry out useful case studies to establish the physics linking atomic gas, through the formation of the molecular medium, directly to star formation across entire galaxy disks.

\paragraph{Acknowledgments}--- FMM carried out part of the research activities described in this paper with contribution of the Next Generation EU funds within the National Recovery and Resilience Plan (PNRR), Mission 4 - Education and Research, Component 2 - From Research to Business (M4C2), Investment Line 3.1 - Strengthening and creation of Research Infrastructures, Project IR0000034 – ``STILES - Strengthening the Italian Leadership in ELT and SKA''


\bibliographystyle{abbrvnat-maxbibnames4}
\bibliography{chapter}

\end{document}